\documentclass[12pt]{article}

\advance \textheight by \topskip
\makeatletter
\@addtoreset{equation}{section}
 
\makeatother

\def\bz{\bar z}
\def\bZ{\bar Z}
\def\R{\mathbb R}

\def\N{\mathbb N}
\def\Z{\mathbb Z}

\def\asl{\mathfrak{sl}}

\def\su{\mathfrak{su}}
\def\h{\mathcal{H}}
\def\cF{\mathcal F}
\def\cT{\mathcal T}
\def\cP{\mathcal P}
\def\cD{\mathcal D}

\def\I{\hbox{{1\hskip -5.8pt 1}\hskip -3.35pt I}}

\def\ket #1{|{#1}\rangle}
\def\d{\partial}
\def\bd{\bar\partial}
\def\mop #1{\mathop{\sf #1}\nolimits}

\usepackage{amsmath,amssymb}

\begin{document}
\title{
\begin{flushright}
\parbox{3.5cm}{
{\small USACH-FM-02/08\\[-3mm]
hep-th/0212117}}
\end{flushright}
\vskip 2cm
{\bf Dolan-Grady Relations and Noncommutative
Quasi-Exactly Solvable Systems}}

\author{{\sf Sergey M. Klishevich${}^{a,b}$}\thanks{
E-mail: sklishev@lauca.usach.cl}
{\sf\ and Mikhail S. Plyushchay${}^{a,b}$}\thanks{
E-mail: mplyushc@lauca.usach.cl}
\\
{\small {\it ${}^a$Departamento de F\'{\i}sica,
Universidad de Santiago de Chile,
Casilla 307, Santiago 2, Chile}}\\
{\small {\it ${}^b$Institute for High Energy Physics,
Protvino, Russia}}}
\date{}

\maketitle

\vskip-1.0cm

\begin{abstract}
 We investigate a $U(1)$ gauge invariant quantum
 mechanical system on a 2D noncommutative space with
 coordinates generating a generalized deformed
 oscillator algebra. The Hamiltonian is taken as a
 quadratic form in gauge covariant derivatives obeying
 the nonlinear Dolan-Grady relations. This restricts
 the structure function of the deformed oscillator
 algebra to a quadratic polynomial. The cases when the
 coordinates form the $\su(2)$ and $\asl(2,\R)$
 algebras are investigated in detail. Reducing the
 Hamiltonian to 1D finite-difference quasi-exactly
 solvable operators, we demonstrate partial
 algebraization of the spectrum of the corresponding
 systems on the  fuzzy sphere and noncommutative
 hyperbolic plane. A completely covariant method based
 on the notion of intrinsic algebra is proposed to deal
 with the spectral problem of such systems.
\end{abstract}

\newpage
\section{Introduction}

The Dolan-Grady relations (DGR) arose 20 years ago in a
very interesting  paper \cite{dg}, where Dolan and Grady
considered
a class of systems described by a Hamiltonian given as a
linear combination of the two operators obeying some
nonlinear relations. The relations, afterward called by
their names, guarantee the existence of an infinite set of
mutually commuting charges which includes the Hamiltonian.
This naturally connects the construction to the integrable
systems \cite{dg,Perk,dav,gehlen,baxter}. A rich
algebraic structure associated with the Dolan-Grady
relations was revealed by Perk \cite{Perk} and Davies
\cite{dav} who showed that the two operators
satisfying DGR generate recursively the infinite-dimensional
Onsager algebra. The latter appeared in a seminal work
\cite{onsager} on exact solution of the planar Ising model.
The set of commuting charges constructed by Dolan and Grady
is linear in generators of the Onsager algebra.

Recently it was observed \cite{susy-pb,d1,nsusy} that
the Dolan-Grady relations arise as necessary conditions
for the anomaly-free quantization of pseudo-classical
systems with nonlinear holomorphic supersymmetry
($n$-HSUSY). In accordance with general algebraic
construction of the $n$-HSUSY \cite{nsusy}, the
Hamiltonian depending on a natural parameter $n$ is a
quadratic form in two mutually conjugate operators
subjected to DGR, whereas the supercharges are (anti)
holomorphic polynomials of the order $n$ in the same
operators. It was shown that the $n$-HSUSY Hamiltonian
belongs to an infinite set of mutually commuting even
operators being quadratic in the Onsager algebra
generators, and that these even integrals together with
the odd supercharges form a polynomial superalgebra of
the order $n$. Realization of this construction in
various physical systems \cite{d1,d2,nRS} revealed an
intimate relation of the $n$-HSUSY with the
quasi-exactly solvable (QES) systems \cite{qes}.

Though the quadratic set of commuting integrals admits
any real and even complex values for the parameter $n$
playing the role of a coupling constant, such an
extension breaks the nonlinear supersymmetry.
Nevertheless, in Ref. \cite{nRS} it was observed that
the quasi exact solvability of the systems associated
with the  quadratic set of integrals of motion roots
not in the $n$-HSUSY  construction itself, but rather
in the Dolan-Grady relations. In particular, it was
shown that the corresponding systems can be
quasi-exactly solvable not only for integer values of
the coupling constant $n$.

{\it In the present paper, exploiting the Dolan-Grady
relations we consider a class of $U(1)$ gauge invariant
quantum mechanical systems on a two-dimensional
noncommutative manifold, and demonstrate the quasi
exact solvability of their spectral problem for any
value of the coupling constant.}

The layout of the paper is as follows. In Section 2 the
basic information on the Dolan-Grady relations and
Onsager algebra necessary for further analysis is
presented in the context of integrable systems,
$n$-HSUSY and low-dimensional matrix models.
Realization of the Dolan-Grady relations in terms of
the generalized deformed oscillator algebra is given in
Section 3. In Section 4 we construct a $U(1)$ gauge
invariant quantum mechanical system with noncommutative
coordinates generating the generalized deformed
oscillator  algebra. In Section 5 we demonstrate that
by a reduction to finite-difference equations, the
spectral problem of the model on the fuzzy sphere and
noncommutative hyperbolic plane can be partially solved
for some values of the coupling constant. In Section 6
we develop an algebraic covariant approach to deal with
the spectral problem, and show that really our
noncommutative system is quasi-exactly solvable for any
value of the coupling constant. The obtained results
are summarized in Section 7. In Appendix we discuss a
deformation of the $\asl(2,\R)$ scheme associated with
the finite-difference quasi-exactly solvable equations.

\section{Dolan-Grady relations}

\subsection{Dolan-Grady relations and Onsager algebra}

The Dolan-Grady relations \cite{dg} can be represented
in the form \cite{d2,nsusy}
\begin{align}\label{dgr}
 \left[Z,\:\left[Z,\:\left[Z,\:\bar Z\right]\right]
 \right]&=
 \omega^2\left[Z,\:\bar Z\right],&
 \left[\bar Z,\:\left[\bar Z,\:
 \left[Z,\:\bar Z\right]\right]\right]&
 =\bar\omega^2\left[Z,\:\bar Z\right].
\end{align}
Here $Z$ and $\bar Z$ are some operators which we shall
call generating elements while $\omega$ and
$\bar\omega$ are some constants. The nonlinear
relations (\ref{dgr}) can be used to construct the
infinite set of mutually commuting operators for some
integrable systems. In the most elegant way they can be
represented in terms of the Onsager algebra
\cite{onsager} spanned by the generators $A_n$ and
$G_n$ having the commutation relations
\begin{align}\label{OA0}
 \left[A_m,\:A_n\right]&=4G_{m-n},&
 \left[G_m,\:A_n\right]&=2A_{n+m}-2A_{n-m},&
 \left[G_m,\:G_n\right]&=0,
\end{align}
where $m,n\in\Z$. In fact, the operators
$A_0\sim\bar Z$ and $A_1\sim Z$ are the generating
elements of the Onsager algebra obeying the Dolan-Grady
relations and all the other operators are defined
recursively by the relations (\ref{OA0}) \cite{dav}.
The values of the constants $\omega$ and $\bar\omega$
can be fixed by rescaling the generators. Then, the
infinite set of the mutually commuting operators
constructed by Dolan and Grady \cite{dg} is represented
as
\begin{equation}\label{lin}
 2J_m=A_m+A_{-m}+\lambda\left(A_{m+1}+A_{1-m}\right),
\end{equation}
where $\lambda\in\R$. The commutativity of the
operators $J_m$ follows directly from the algebra
(\ref{OA0}). Now, treating the operator
$J_0=A_0+\lambda A_1$ as a Hamiltonian, one obtains,
generally, an integrable system with infinite number of
conserved commuting charges. The charges of the
form (\ref{lin}) appear in various models, such as
the Baxter eight-vertex \cite{baxter}, the
two-dimensional Ising \cite{onsager} and $Z_N$ spin
models \cite{gehlen}. The power of the
algebraic formulation (\ref{OA0}), (\ref{lin}) is that
it does not refer to the number of dimensions or to the
nature of the space-time manifold which can be
lattice, continuum or loop space.

\subsection{Structure of $n$-HSUSY}

In the systems associated with the linear set of
integrals (\ref{lin}) the operators $Z$ (${}\sim A_1$)
and $\bar Z$ (${}\sim A_0$) are supposed to be
Hermitian. In Ref. \cite{nsusy} it was shown that the
set (\ref{lin}) is not a unique set of mutually
commuting operators which can be constructed in terms
of generators of the Onsager algebra. There exists a
set of commuting operators which are quadratic in the
generators of the algebra (\ref{jmn}). Such commuting
operators naturally appear in the scheme of the
$n$-HSUSY \cite{d1,d2,nsusy,nRS}, which we discuss
shortly below. Note, that for the systems associated
with the quadratic set of commuting operators, the $Z$,
$\bar Z$ and the $\omega$, $\bar\omega$ are usually
supposed to be mutually conjugate operators and complex
constants, respectively.

The Hamiltonian of the system associated with the
quadratic set of commuting charges
(see Eq. (\ref{jmn}) for their explicit form) reads as
\begin{equation}\label{H}
 H_\lambda=\frac 14\left\{\bar Z,\;Z\right\}
   -\frac\lambda 4\left[Z,\;\bar Z\right],
\end{equation}
where a real parameter $\lambda$ serves as a coupling
constant, while mutually conjugate operators $Z$ and
$\bar Z$ are supposed to obey the nonlinear Dolan-Grady
relations (\ref{dgr})\footnote{In principle, one can
also deal with Hermitian operators $Z$ and $\bar Z$.
Then the Hamiltonian (and the whole set of the
integrals (\ref{jmn})) is Hermitian if the coupling
constant is pure imaginary \cite{nsusy}. However, here
we will not discuss such systems.}.

When the coupling constant is integer, $\lambda=\pm n$,
$n\in\Z_+$, the system (\ref{H}) can be extended to the
supersymmetric system with the Hamiltonian
$\mathcal H_n=H_n\sigma_-\sigma_++H_{-n}\sigma_+\sigma
_-$, $\sigma_\pm=\frac 12(\sigma_1\pm i\sigma_2)$. In
this case the Dolan-Grady relations (\ref{dgr})
guarantee the existence of the odd integrals of motion
(supercharges)
\begin{align}
 Q&=\sigma_+\prod_{k=0}^{n-1}
 \left(Z-\frac\omega 2(2k-n+1)\right),&
 \bar Q&=\sigma_-\prod_{k=0}^{n-1}
 \left(\bZ-\frac{\bar\omega}2(2k-n+1)\right).
\end{align}
The supercharges generate a nonlinear superalgebra of
the order $n$. Its structure and further details of the
algebraic construction underlying the nonlinear
holomorphic supersymmetry can be found in
Ref.~\cite{nsusy}.

The advantage of the supersymmetric formulation is that
the zero modes of the supercharge $Q$ (or $\bar Q$)
form an invariant subspace with respect to the action
of the Hamiltonian $H_n$ (or $H_{-n}$). Furthermore, as
a rule the zero modes are easier calculable than the
eigenfunctions of the Hamiltonian due to the factorized
structure of the supercharges. When the number of the
zero modes is finite, then the Hamiltonian is
quasi-exactly solvable operator. The information on the
functional form of the zero modes is very helpful for
solving the corresponding spectral problem.

Although the system (\ref{H}) formally admits an
infinite number of integrals, nevertheless, for any
given realization of the generating elements in a
system with finite number of degrees of freedom this
infinite set is reduced to a finite set of independent
integrals. Such systems being not integrable in general
case, however, reveal a profound connection with
quasi-exactly solvable systems \cite{nsusy}. We shall
return to this aspect below.

\subsection{Matrix models}

Here we show that the Dolan-Grady relations
can be revealed in some low dimensional matrix models.

So, let us
consider a matrix
model given by
the action
\begin{equation}\label{S2}
  S ={}-\mop{Tr}
 \left(\frac 12\left[X_\mu,\,X_\nu\right]
 \left[X^\mu,\,X^\nu\right]
 + R_{\mu\nu}X^\mu X^\nu\right),
\end{equation}
where $X_\mu$ are Hermitian operators on a Hilbert
space and $R_{\mu\nu}$ are $c$-numbers. The last term
can be treated as describing the interaction of the
vector field $X_\mu$ with an external symmetric field
$R_{\mu\nu}$. One can also think that its diagonal part
contains the mass term of the vector field. The matrix
models of the form (\ref{S2}) are widely used in the
context of string theories. In the case of positively
definite metric the action (\ref{S2}) corresponds to
the potential part of the BFSS model \cite{BFSS} and
defines static (vacuum) configurations of the theory.
On the other hand, for the Lorentz metric the action
can be interpreted as an effective theory of the
reduced model of Yang-Mills theory \cite{IKKT} with the
specific interaction. Actually, action (\ref{S2})
represents a slightly modified massive matrix
model~\cite{mmm}.

The action (\ref{S2}) leads to the equations of motion
\begin{equation}\label{YM}
  [X^\nu,\,[X_\mu,\,X_\nu]] + R_{\mu\nu}X^\nu=0.
\end{equation}
In the two-dimensional case Eq. (\ref{YM}) can be
represented in the form
\begin{align}
 [X_-,\,[X_-,\,X_+]] + R_{+-}X_- + R_{--}X_+&=0,
 \notag\\[-3mm]\label{para}\\[-3mm]\notag
 [X_+,\,[X_+,\,X_-]] + R_{+-}X_+ + R_{++}X_-&=0,
\end{align}
where $X_\pm=\frac{1}{\sqrt{2}}(X_1\pm X_2)$ or
$X_\pm=\frac{1}{\sqrt{2}}(X_1\pm iX_2)$ for the
Lorentz  or Euclidean metric, respectively. In the
first
case $R_{++}$ and $R_{--}$ are real constants, while
in the
latter
$R_{--}^*=R_{++}$. It is easy to see that any solution
of
Eq. (\ref{para}) also obeys the Dolan-Grady relations
(\ref{dgr}) with the
identification:
\begin{align*}
 X_-&\sim Z,&
 X_+&\sim \bar Z,&
 R_{--}&\sim -\omega^2,&
 R_{++}&\sim -\bar\omega^2.
\end{align*}
Moreover, note that in the case of positively definite
metric and
when $R_{++}=R_{--}=0$, the equations
(\ref{para}) coincide with the basic commutations
relations for the single-mode parafermion \cite{kam}.

Now let us consider the following modification of the
three-dimensional matrix model \cite{YM+CS}
\begin{equation}\label{S3}
  S ={}- \mop{Tr}
 \left(\frac 12[X_j,\,X_k]^2
 -\frac 43i\alpha\epsilon_{jkl}X_jX_kX_l
 +R_{jk}X_jX_k\right),
\end{equation}
where $\alpha$ and $R_{jk}$ are real numbers and we
imply
positively definite  metric. This model can be treated
as
the matrix model corresponding to the usual Yang-Mills
theory with Chern-Simon term. As before, the quadratic
term
in action (\ref{S3}) represents a specific interaction
with an external symmetric field. The equations of
motion
are
\begin{equation}\label{eq3}
 [[X_k,\,X_j],\,X_k]-i\alpha\epsilon_{jkl}[X_k,\,X_l]
 +R_{jk}X_k=0.
\end{equation}
Imposing the condition
\begin{equation}
 [X_0,\,X_\pm]=\pm\beta X_\pm
\label{grad}
\end{equation}
with $\beta\in\R$ on the components
$X_0$, $X_{\pm}$, the equation for $j=0$ is satisfied
identically for
$\beta =\alpha$, $R_{0i} = 0$.
The equations of motion for $X_\pm$ have the
form (\ref{para}) with the substitution
$R_{+-}\to R_{+-}+\alpha$.
Therefore, the Dolan-Grady relations arise in this
model as
well. Again, like in model (\ref{S2}), for
$R_{\pm\pm}=0$ the equations of motion for
the components $X_\pm$ take the form of
basic commutation relations of
the single-mode parafermion.

\section{Generalized deformed oscillator}

\subsection{Main definitions}

Another basic ingredient we shall use
in what follows is the generalized
deformed oscillator (GDO). Therefore, in this
subsection we
present a brief description of the construction
(see also Refs. \cite{gdo,gdosu}).

The generalized deformed oscillator is
defined by the algebra generated by the set
of operators
$\{\I,z,\bz,N\}$ satisfying the relations
\begin{align}\label{gen}
[N,\,z]&=-\theta z,&
[N,\,\bz]&=\theta\bz,&
 \bz z&=F(N),& z\bz&=F(N+\theta),
\end{align}
where $N$ is the number operator and the structure
function
$F(.)$ is an analytic function with the properties
$F(0)=0$ and $F(n\theta)>0$, $n=1,\ldots$. The
structure
function $F(.)$ is characteristic to the deformation
scheme.
Here we have introduced a positive real parameter $
\theta$,
which will serve later on as a measure of
noncommutativity.

The operators $z$, $\bz$ obey the commutation relation
\begin{equation}\label{com}
 [z,\,\bz]=\theta\Delta_+F(N).
\end{equation}
By symbols $\Delta_\pm$ we denote the
forward and backward difference derivatives
\begin{align}\label{dd}
 \Delta_+f(N)&=\frac{f(N+\theta)-f(N)}\theta,&
 \Delta_-f(N)&=\frac{f(N)-f(N-\theta)}\theta.
\end{align}

The generalized deformed oscillator algebra can be
naturally
represented on the Fock space of the eigenstates of the
number
operator $N$,
$N\ket{n}=n\theta\ket{n}$,
$\langle n|m\rangle=\delta_ {nm}$, $n=1,\ldots$, with
the
vacuum state defined as $z\ket{ 0}=0$. The eigenstates
are
given by
\begin{align}\label{Hilb}
 \ket{n}&=\frac 1{\sqrt{F(n\theta)!}}\bz^n\ket{0},&
 F(n\theta)!&\equiv \prod_{k=1}^nF(k\theta).&
\end{align}
The operators $z$ and $\bz$ are the annihilation and
creation operators of the oscillator algebra,
\begin{align}\label{ancr}
 z\ket{n}&=\sqrt{F(n\theta)}\ket{n-1},&
 \bz\ket{n}&=\sqrt{F((n+1)\theta)}\ket{n+1}.
\end{align}
In what follows we shall denote the Fock space defined
by
(\ref{Hilb}), (\ref{ancr}) as $\h_F$. On this space
the
operators $z$ and $\bz$ are mutually adjoint,
$\bz=z^{\dag}$.

Note that when the characteristic function obeys the
condition
\begin{equation}\label{p+1}
 F((p+1)\theta)=0
\end{equation}
for some $p\in\N$, the creation-annihilation operators
satisfy the nilpotency relations
$z^{p+1}=\bz^{p+1}=0$. This
means that the corresponding representation is
$(p+1)$-dimensional and we have a parafermionic type
system
of the order $p$.

\subsection{The Dolan-Grady relations in terms of GDO}

The simplest realization of the Dolan-Grady relations
(\ref{dgr}) can be obtained in terms of the
generalized
deformed oscillator:
\begin{equation}\label{Z=z}
 Z=z,\hskip 2.5cm \bZ=\bz.
\end{equation}
As a consequence of the natural grading of the GDO
algebra
defined by the operator $N$, for this representation
of the
generating elements $Z$ and $\bar Z$ it is possible to
realize the
Dolan-Grady relations only in its contracted form,
$\omega=0$
\cite{nsusy},
\begin{align}\label{cdgr}
 \left[Z,\:\left[Z,\:\left[Z,\:\bZ\right]\right]
 \right]&=0,&
 \left[\bZ,\:\left[\bZ,\:\left[Z,\:\bZ\right]
  \right]\right]&=0.
\end{align}
On the other hand,  these relations lead to the
restriction
on the structure function in the form of the
finite-difference equation:
\begin{equation}\label{F}
 \Delta_+^3F(N)=0.
\end{equation}
Since the structure function has to vanish at zero,
the
general solution of the equation (\ref{F}) is
\begin{equation}\label{Fsol}
 F(N)=c_1N^2 + c_0\theta N,
\end{equation}
where {\it ad interim} $c_0,c_1\in\R$. Three different
cases corresponding to the solution (\ref{Fsol})
should be distinguished:
\begin{enumerate}
 \item
 The case $c_1=0$, $c_0\ne 0$ corresponds to the usual
 oscillator given by the Heisenberg algebra. The
 nonlinear supersymmetry for such a  system was
 discussed in Ref. \cite{susy-pb}.

 \item
 For $c_1>0$ the polynomial (\ref{Fsol}) is a structure
 function when the parameters obey the inequality
 $c_1+c_0>0$. Then the generalized oscillator realizes
 the half-bounded infinite-dimensional discrete series
 of the unitary representations of the $\asl(2,R)$
 algebra. The generators are, correspondingly,
 $L_0= \theta\Delta_+F(N)$, $L_-=z$, $L_+=\bz$, or
 $L_0=-\theta\Delta_+F(N)$, $L_-=\bz$, $L_+=z$ for the
 series $\cD^+_\alpha$ or $\cD^-_\alpha$ \cite{bargm}
 with $\alpha=(c_0+c_1)/(2c_1)$.

 \item
 In the case $c_1<0$, for the solution (\ref{Fsol}) to
 be a structure function one has to impose the
 additional condition (\ref{p+1}). This leads to
 quantization of one of the parameters,
 $c_0=-(p+1)c_1 $, and results in the
 $(p+1)$-dimensional representation of the $\su(2)$
 algebra with the generators $L_-=z$, $L_ +=\bz$ and
 $L_3=\theta\Delta_+ F(N)$. This is the single-mode
 parafermion case.
\end{enumerate}

\section{Noncommutative quantum mechanics and DGR}

Let us consider a 2D quantum mechanical system with
noncommutative complex-like coordinates $z$, $\bz$
obeying
the GDO algebra (\ref{gen}) realized on the Hilbert
space
$\h_F$.
The operator-valued functions $\Psi(\bz,z)$ on
$\h_F$ with a finite norm with respect to the scalar
product
\begin{equation}\label{SP}
 \left(\Phi,\,\Psi\right)\equiv
 {\sf Tr}_{\h_F}\bigl(\Phi^{\dag}(\bz,z)\Psi(\bz,z)
 \bigr)
\end{equation}
will be treated by us as
the states of the noncommutative system.
Here the symbols ${\sf Tr}_{\h_F}$ and ${}^{\dag}$
denote
the usual trace and Hermitian conjugation operations
on the
space $\h_F$ (\ref{Hilb}). The operator-valued
functions
with the finite norm themselves form a Hilbert space,
which
we shall denote as $\hat\h_F$.

The corresponding derivative operators on the space
$\hat\h_F$ can be defined as
\begin{equation}\label{der}
 \d\equiv -\theta^{-1}\mop{ad}\bz,\hskip 2cm
 \bd\equiv\theta^{-1}\mop{ad}z.
\end{equation}
These operators possess the usual conjugation property
$\d^{\dag}=-\bd$ with respect to the scalar product
(\ref{SP}). It is worth noting that the derivative
operators
do not commute if the structure function $F(N)$ is
different
from a linear one. Indeed,
$[\d,\,\bd]=\theta^{-1}\mop{ad}\Delta_+ F(N)$.

Let us introduce a $U(1)$ gauge interaction associated
with
an external magnetic field. The noncommutative
system admits the following two types of gauge
transformations:
\begin{align}\label{transV}
 \Psi'(\bz,z)&=U(\bz,z)\Psi(\bz,z),&
 \Psi'{}^{\dag}(\bz,z)&=\Psi^{\dag}(\bz,z)U^{\dag}(
 \bz,z),
\end{align}
or
\begin{equation}\label{transT}
 \Psi'(\bz,z)=U(\bz,z)\Psi(\bz,z)U^{\dag}(\bz,z)
\end{equation}
with $U(\bz,z)$ being a unitary operator. The
scalar product (\ref{SP}) is invariant with respect to
the
both transformations. In the commutative
limit, $\theta\to 0$, the transformation
(\ref{transV}) is reduced to the usual $U(1)$
gauge
transformation while (\ref{transT})
becomes trivial.

We realize the generating elements
from (\ref{H})
as the gauge
covariant derivatives
\begin{align}\label{Znc}
 Z&=\d+\rho\bigl[A(\bz,z)\bigr],&
 \bZ&={}-\bd+\rho\bigl[\bar A(\bz,z)\bigr],
\end{align}
where the operator $\rho[{\cal O}]$,
${\cal O}\in\{A(\bz,z),\bar A(\bz,z)\}$, acts on
$\Psi,\Psi^{\dag}$ as
$$
 \rho[{\cal O}]\Psi={\cal O}\Psi, \hskip 2cm
 \rho[{\cal O}]\Psi^{\dag}=-\Psi^{\dag}{\cal O}
$$
manifesting the fact that the states $\Psi$ and $\Psi^
\dag$
carry the charges of the opposite sign.
The operator-valued functions $A$ and  $\bar
A=A^{\dag}$
are the components of the noncommutative vector
gauge potential
corresponding to the case of stationary magnetic
field, and
we assume that the gauge coupling constant (``electric
charge'') and the mass parameter are equal to unit,
$e=m=1$. Under the both gauge transformations
(\ref{transV}) and (\ref{transT}) the components of
the noncommutative vector potential are transformed in
the usual way
\begin{align*}
 A'&=UAU^{\dag}+U\d U^{\dag},&
 \bar A'&=U\bar AU^{\dag}-U\bd U^{\dag}.
\end{align*}

The commutator of the generating elements has the form
\begin{align}\label{comZ}
 \mathcal B&\equiv[Z,\,\bZ]=
  {}-\theta^{-1}\mop{ad}\Delta_+ F(N) + B(\bz,z).
\end{align}
We will treat
\begin{equation}\label{B}
 B(\bz,z)\equiv\d\bar A + \bd A +\left[A,\,\bar A
 \right]
\end{equation}
as an external quasi-magnetic field since in the
commutative limit the commutator in (\ref{B})
disappears and it turns into a true magnetic field.
Nevertheless, in the noncommutative case the quantity
(\ref{B}) transforms as a connection under the gauge
transformations. Actually, if one considers the
operator $N$ as the third dependent coordinate, then
the quantity (\ref{B}) is the $N$-th component of the
gauge connection.

Let us suppose that the magnetic field depends on $N$
only. This corresponds to an axially symmetric field in
the commutative limit. The following choice of the
gauge,
\begin{equation}\label{gauge}
 A(\bz,z)=\bz f(N), \hskip 2cm \bar A(\bz,z)=f(N)z,
\end{equation}
where $f(.)$ is a real function, guarantees such a
dependence of the magnetic field. Indeed, in the gauge
(\ref{gauge}) the quasi-magnetic field (\ref{B})
acquires
the form
\begin{equation}\label{BN}
 B(N)=\Delta_+\bigl(
 F(N)f(N-\theta)
  \left(2-\theta f(N-\theta)\right)\bigr).
\end{equation}

Now let us analyze the restrictions which follow from the
Dolan-Grady relations. In the representation
(\ref{Znc})
the l.h.s. of Eq. (\ref{dgr}) can be rewritten as
follows
\begin{gather}
l.h.s.(\ref{dgr}) =
{}-\theta^{-1}\mop{ad}\bigl(\bz^2\Delta_+^3F(N)\bigr)
   + \bz^2\left(1-\theta f(N+\theta)\right)
     \left(1-\theta f(N)\right)\Delta_+^2B(N)
   \notag\\[-2.5mm]\label{cdgrZ}\\[-2.5mm]\notag
 {}+ \bz^2\bigl(f(N)+f(N+\theta)(1-\theta f(N))\bigr)
    \Delta_+^3F(N).
\end{gather}
The comparison of the first terms in (\ref{comZ}) and
(\ref{cdgrZ}) leads to the conclusion that the
Dolan-Grady relations (\ref{dgr}) with $\omega\ne 0$
can be satisfied only when the structure function of
the GDO algebra is linear. On the other hand, for
$\omega=0$ the expression (\ref{cdgrZ}) vanishes only
when the structure function obeys the equation
(\ref{F}). This leads to the following constraint on
the quasi-magnetic field:
\begin{equation}\label{eqB}
 \Delta_+^2B(N)=0.
\end{equation}
In other words, the quasi-magnetic field has to be
at most linear in $N$. The function $f(N)$ appeared in
the gauge (\ref{gauge}) can be found from (\ref{BN})
taking into account the constraint (\ref{eqB}).

The classification of the GDO algebra with the
quadratic
structure function (\ref{Fsol}) was presented in the
previous section. In accordance with it, the
noncommutative
coordinates $z$, $\bz$ and the operator $N$ form the
$\su(2)$,
$\asl(2,\R)$, or the Heisenberg algebras. Therefore,
the
noncommutative system given by the Hamiltonian
(\ref{H}) in
the representation (\ref{Znc}) can be treated,
correspondingly, as the system on the fuzzy sphere $S^
2_
\theta$, on the noncommutative hyperbolic plane
$H_\theta^2$,
or on the noncommutative plane $\R^2_\theta$.

Note that since the structure function is quadratic,
the simplest solution for $f(N)$ is
\begin{equation}\label{f}
 f(N)=f=const.
\end{equation}
Indeed, one can verify that in this case the
quasi-magnetic field (\ref{B}) is a linear function,
\begin{equation}\label{B-N}
 B(N)=f(2-\theta f)(2c_1N+(c_0+c_1)\theta),
\end{equation}
and hence, it obeys the equation (\ref{eqB}). Below we
shall
consider only the solution (\ref{f}), (\ref{B-N}).
Later on
we shall show that for the given solution the
expression
(\ref{B-N}) coincides with a covariant definition of
the
magnetic field in the gauge (\ref{gauge}).

{}From the explicit form of the magnetic field
(\ref{B-N})
it follows that the parameter $f$ has two special
values $f=
0$ and $f=2\theta^{-1}$. The first case is rather
trivial.
It corresponds to the model without gauge interaction
(a free model (\ref{H})
on the corresponding noncommutative space).
In
the second case the magnetic field (\ref{B-N})
vanishes.
Nevertheless, the gauge interaction still is present
in the
model since the corresponding connection
(\ref{gauge}) with $f\ne 0$  is not trivial.
This resembles the famous Aharonov-Bohm
effect but its origin roots here
in the noncommutativity of the
configuration space. As we shall see, there is
also another
critical value of the parameter, $f=\theta^{-1}$. For
this
value the
quasi-magnetic field is nontrivial while
the system is degenerate.
Such a  case, in turn,  resembles the situation with
existence of the critical value of the  constant external
magnetic field being coherent with the value of
the parameter of noncommutativity
(see, e.g. \cite{noncomplane,nair}).

In the next section we shall discuss the spectral
problem
for the noncommutative system (\ref{H}) on the
fuzzy sphere and noncommutative hyperbolic plane.

\section{Reduction to finite-difference QES systems}

Here we show that the noncommutative
system given by the Hamiltonian (\ref{H}) is
quasi-exactly solvable for some values of the coupling
constant. It is demonstrated by a reduction of the
Hamiltonian to one-dimensional finite-difference QES
operators.

{}First of all, one can notice that in the gauge
(\ref{gauge}) the operator $\mop{ad}N$ commutes with
the
Hamiltonian. One can refer this property to the
existence in
the system of an ``axial'' symmetry. Therefore, the
eigenstates of the Hamiltonian can be represented in
the
form
\begin{equation}\label{modes}
 \Psi_L(\bz,z)=\bz^m\psi(N),
  \hskip 1cm\text{or}\hskip 1cm
 \Psi_R(\bz,z)=\psi(N)z^m,
\end{equation}
where $m\in\Z_+$. Evidently, these operator functions
are
eigenstates of the operator $\mop{ad}N$.
We shall call  the functions
$\Psi_L$ and $\Psi_R$
the left and right modes.

Now it is clear enough that the corresponding 2D
spectral
equation with the eigenstates of the form
(\ref{modes}) can
be reduced to a one-dimensional problem. But unlike
the commutative case, here the corresponding 1D system
is
represented by a finite-difference equation. One notes
that such a situation is typical for any
noncommutative
system with the ``axial'' symmetry.

The Hamiltonian (\ref{H}) is associated with the
nonlinear
holomorphic supersymmetry for integer values of the
coupling
constant. The zero modes of the supercharge $Q$ form
an
invariant subspace of the Hamiltonian. We will use
their
functional form as an anzatz for investigating the
spectral
problem with arbitrary coupling constant.

Let us temporarily set $\lambda=n\in\N$. For the left
mode
wave functions represented in the factorized form
$$
 \Psi_L(\bz,z)=\bz^m\varphi(N)\phi(N),
$$
with $\phi(N)$ obeying the equation
\begin{equation}\label{phi_f}
 (\Delta_++f)\phi(N)=0,
\end{equation}
the zero mode equation $Z^n\Psi_L=0$ is reduced
to
$$
 (1-\theta f)^n\Delta_+^n\varphi(N)=0.
$$
For $\theta f\ne 1$ this equation gives the polynomial
solution $\varphi(N)=P_{n-1}(N)$, where $P_k(.)$
denotes a
polynomial of the $k$-th degree. For $\theta f=1$
there are
no zero modes at all, i.e. there are no bounded
(normalized) states in the system.

The existence of the zero modes means that the
Hamiltonian
is quasi-exactly solvable when the coupling constant
$\lambda$
is a nonnegative integer. Now we investigate the
question on existence of the zero modes for
an arbitrary value of the coupling constant.
For the left modes the Hamiltonian
can be reduced,
$
 H_\lambda\Psi_L(\bz,z)\to H_{\lambda,m}^L\psi(N),
$
to the one-dimensional finite-difference operator
\begin{align}\label{Hl}
 H_{\lambda,m}^L&=
  (\theta f-1)\bigl(F(N)\,\Delta_+\Delta_-
    + (m+1){\cal F}(N)\,\Delta_+\bigr) + f^2F(N)
    \notag\\[-3mm]\\[-3mm]\notag&{}
     + \frac 12f\bigl((2m+\lambda+1)\theta f
     -2(m+\lambda)\bigr)\Delta_+F(N) + const
\end{align}
with the function ${\cal F}(N)$ defined by
\begin{align}\label{cF}
 {\cal F}(N)&=\Delta_+F(N)+\frac{m\theta}2\Delta_+^2F(
 N).
\end{align}
For $\theta f=1$ the reduced
Hamiltonian (\ref{Hl}) is a multiplicative operator.
The generating elements  of the form (\ref{Znc}) are
multiplicative operators as well, $Z\Psi\sim\Psi\bz$,
$\bar Z\Psi
\sim\Psi z$. Actually, this representation of the
Dolan-Grady relations becomes equivalent to the
representation
(\ref{Z=z}). However, the Hamiltonian (\ref{H}) acting
on
the Hilbert space $\hat\h_F$ has a trivial dynamics in
this
case.
Therefore, later on we shall always assume
$\theta f\ne 1$.

The leading term of the ``potential'' in (\ref{Hl}) is
proportional to the structure function $F(N)$. We
represent
the function $\psi(N)$ in the factorized form
\begin{equation}\label{psi}
 \psi(N)=\varphi(N)\phi(N),
\end{equation}
where the function $\phi(N)$
is supposed to
obey the equation
\begin{equation}\label{phi_g}
  (\Delta_+ + \gamma)\phi(N)=0
\end{equation}
with a real parameter $\gamma$. This equation is a
slight modification of Eq. (\ref{phi_f}).
The factorization (\ref{psi}) allows us to discard the
leading term in the equation for $\varphi(N)$ when the
constant $\gamma$ obeys the algebraic equation
\begin{equation}\label{g_eq}
 \frac{\gamma^2}{1-\theta\gamma}=\frac{f^2}{1-\theta
 f},
\end{equation}
that gives a possibility to look for the solutions
$\varphi(N)$ in a polynomial form.

The first evident solution of the equation
(\ref{g_eq})
\begin{equation}\label{sol1}
 \gamma=f
\end{equation}
corresponds to the zero modes found above. Indeed, the
reduction
$$
 H_{\lambda,m}^L\psi(N)\to H_{\lambda,m}'\varphi(N)
$$
results in the one-dimensional finite-difference
operator
\begin{equation}\label{Hj=n}
 H_{\lambda,m}'={}-F(N)\,\Delta_+\Delta_-
    - (m+1)(1-\theta f)^2\cF(N)\,\Delta_+
    + f(2-\theta f)\,\cT_\lambda + const.
\end{equation}
Here, the function $\cF(N)$ is defined by (\ref{cF}),
while
the operator $\cT_\lambda$ is given by
\begin{equation}\label{cT}
 \cT_\lambda=F(N)\Delta_+ - \lambda\Delta_+F(N).
\end{equation}

Let $\cP_k(x)$, $k\in\Z_+$, denotes the space of all
the
polynomials in $x$ up to the order $k$. On such spaces
the
finite-difference derivatives act similarly to the
usual
derivative, $\Delta_\pm:\cP_k(N)\to\cP_{k-1}(N)$.
Using this fact and that the function $F(N)$ is
quadratic
while $\cF(N)$ is linear, one can notice that the
spaces
$\cP_k(N)$ are invariant subspaces of
the operator (\ref{Hj=n}) except for the third term.
On
the other hand,
the space $\cP_{n}(N)$ is an
invariant subspace of the operator $\cT_\lambda$
for any $\lambda=n\in\Z_+$.
As a result, for nonnegative integer values
of the coupling constant
the operator (\ref{Hj=n}) is
quasi-exactly solvable \cite{qes}.

The second solution of the equation (\ref{g_eq}) is
\begin{equation}\label{sol2}
 \gamma=\frac f{\theta f-1}.
\end{equation}
As a consequence, the reduced Hamiltonian acting on
$\varphi(N)$ reads as
\begin{align}\label{Hj2}
 H_{\lambda,m}''&=-(1-\theta f)^2
 \bigr(F(N)\,\Delta_+\Delta_- + (m+1)\cF(N)\,\Delta_+
 \bigl)
    - f(2-\theta f)\,\cT_{-\lambda-2m} + const.
\end{align}
Therefore, this operator is quasi-exactly solvable if
one
imposes the condition $\lambda+2m\in\Z_-$ on the
coupling
constant.

Now we pass over to the discussion of the right modes.
Unfortunately, in this case we failed to find the
zero modes of the supercharge $Q$ in a closed form.
Therefore, we shall investigate the conjectural
partial
algebraization of this part of the
spectrum from viewpoint of
the reduction of the Hamiltonian.

For the right modes (\ref{modes}) the 2D Hamiltonian
is
reduced to the finite-difference operator
\begin{align}
 H_{\lambda,m}^R&={}(\theta f-1)
  \bigl(F(N)\,\Delta_+\Delta_- +
  (m+1)\,\cF(N)\,\Delta_+\bigr) + f^2F(N)
    \notag\\[-3mm]\label{Hr}\\[-3mm]\notag&
     + \frac 12f\bigl((\lambda+1)\theta f
    +2(m-\lambda)\bigr)\Delta_+F(N) + const.
\end{align}
This Hamiltonian can be obtained from (\ref{Hl}) by
means of the formal change $\lambda\to\lambda-2m$.
Using this property, one can reproduce all the
corresponding results for the right modes. The
factorization (\ref{psi}) with the conditions
(\ref{phi_g}) and (\ref{g_eq}) for the solution
(\ref{sol1}) leads to the finite-difference operator
containing the term proportional to the operator
$\cT_{\lambda-2m}$. Hence, for $\lambda-2m\in\Z_+$ the
polynomial space $\cP_{\lambda-2m}(N)$ is an invariant
subspace of the obtained operator, i.e. it is QES
operator. For the second solution (\ref{sol2}) the
reduced Hamiltonian contains $\cT_{-\lambda}$, and so,
it is QES operator for non-positive integer values of
the coupling constant.

All the reduced finite-difference operators,
(\ref{Hj=n}), (\ref{Hj2}) and those for the right
modes, can be represented as QES operators of the form
(\ref{gQES}) with the following coefficients:
\smallskip\par\noindent
1. The solution (\ref{sol1}):
 \begin{gather*}
  a_1 = 0,\qquad a_2 = 0, \qquad
  b_+^+ ={} -\frac{c_1}\theta\,(1-\theta f)^2, \qquad
  b_0^+ = c_0,\qquad
  b_-^+ = 0,\qquad
  b_+^- = \frac{c_1}\theta,
  \\
  b_0^- ={}-\left(c_0 + 2(m+1)c_1\right)(1-\theta
  f)^2,
  \qquad
  b_-^- ={}-\theta(m+1)\left(c_0 + (m+1)c_1\right)
    (1-\theta f)^2,
 \end{gather*}
 where $\mathsf N=\lambda$ or $\mathsf N=\lambda-2m$
 for the left or right modes, respectively;
\smallskip\par\noindent
 2. The solution (\ref{sol2}):
 \begin{gather*}
  a_1 = 0,\qquad  a_2 = 0, \qquad
  b_+^+ ={} -\frac{c_1}\theta, \qquad
  b_0^+ = c_0\,(1-\theta f)^2,\qquad
  b_-^+ = 0,
  \\
  b_+^- = \frac{c_1}\theta\,(1-\theta f)^2,\qquad
  b_0^- ={}-c_0 - 2(m+1)c_1, \qquad
  b_-^- ={}-\theta(m+1)\left(c_0 + (m+1)c_1\right),
\end{gather*}
 where $\mathsf N={}-\lambda-2m$ or
 $\mathsf N=-\lambda$
 for the left or right modes, respectively.
\smallskip

It is worth noting that for $\theta f=2$ the both
operators
(\ref{Hj=n}) and (\ref{Hj2}) are exactly solvable
since the
term proportional to $\cT_\lambda$ vanishes.

Let us briefly discuss
a normalizability condition for the found solutions.
Since the function $\varphi(N)$ is a polynomial,
the normalizability has to be provided by the factor
$\phi(N)$. The solution of the equation (\ref{phi_g})
has
the form $\phi(N)\sim(1-\theta\gamma)^N$. Therefore,
the norm of the corresponding state $\Psi\in\hat\h_F$
can be
represented as
\begin{align}\label{norm}
 \left\|\Psi\right\|_{\hat\h_F}&=
  \mop{Tr}_{\h_F}(\Psi^{\dag}\Psi)=C_{\cal N}
  \sum_{n=1}^{\infty}n^K(1-\theta\gamma)^{2n} +
  \ldots.
\end{align}
Here the dots denote less divergent terms in the sense
that if the first term in (\ref{norm}) is convergent,
then the others are convergent as well. The factor
$C_{\cal N}$ is related to the normalization constant
and the integer parameter $K$ is defined by the type of
solution. The requirement of convergence of the series
leads to the condition
\begin{align}\label{norm1}
 |1-\theta\gamma|<1.
\end{align}
Therefore, we have to impose the following restrictions
on the values of the parameter $f$: $\theta f\in(0,2)$
for the solution (\ref{sol1}), and
$\theta f\in(-\infty,0)\cup(2,\infty)$ for the solution
(\ref{sol2}).

Of course, in the case of the fuzzy sphere the norm
(\ref{norm}) is finite because of the finitness of the
corresponding Hilbert space, and no
additional condition appears for it.

\section{Intrinsic algebra and quasi exact solvability}

In this section we show that
the noncommutative system under consideration really
is quasi-exactly solvable for any value of the coupling
constant. The demonstration is realized in terms
of an intrinsic algebra.

\subsection{Intrinsic algebra}
In general, the operators $Z_0\equiv Z$ and
$\bar Z_0 \equiv \bar Z$ together with the contracted
Dolan-Grady relations (\ref{cdgr}) recursively generate
the infinite-dimensional contracted Onsager algebra
\cite{nsusy}:
\begin{align}
 \left[Z_k,\:\bar Z_l\right]&=B_{k+l+1},&
 \left[Z_k,\:B_l\right]&=Z_{k+l},&
 \left[B_k,\:\bar Z_l\right]&=\bar Z_{k+l},
 \notag\\[-2.5mm]\label{cOA}\\[-2.5mm]\notag
 \left[Z_k,\:Z_l\right]&=0,&
 \left[\bar Z_k,\:\bar Z_l\right]&=0,&
 \left[B_k,\:B_l\right]&=0,
\end{align}
where $k,l\in\Z_+$ and $B_0\equiv 0$ is implied. This
algebra can be extended discarding the last condition,
i. e. by treating $B_0$ as a nontrivial generator. Then
the extended algebra will be generated by the set $Z$,
$\bZ$ and $B_0$. The element $B_0$ splits up the
contracted Onsager algebra into the three subalgebras
of the grade $+1$, $0$ and $-1$, given,
correspondingly, by the span of $\bar Z_m$, $B_m$ and
$Z_m$. In other words, the generator $B_0$ is a grading
element which provides the extended algebra with the
triangular decomposition. This grading is different
from that considered in Ref. \cite{nsusy}.

The algebra (\ref{cOA}) admits the infinite set of the
commuting quadratic charges
\begin{equation}\label{jmn}
 \mathsf J_\lambda^k=\frac 12\sum_{p=1}^k
 \left\{\bar Z_{p-1},\,Z_{k-p}\right\}
  - \frac 12\sum_{p=1}^{k-1}B_pB_{k-p} - \frac\lambda
  2B_k
\end{equation}
with $k\in\Z_+$, which contains the Hamiltonian
(\ref{H}), $\mathsf J^1_\lambda=2H_\lambda$, and the
grading
operator, $\mathsf J^0_\lambda\sim B_0$.

In the gauge (\ref{gauge}) the generating elements of
the form (\ref{Znc}) read as
\begin{align*}
  Z&={}-\theta^{-1}\mop{ad}\bar z+f\rho[\bar z],&
 \bar Z=&{}-\theta^{-1}\mop{ad}z+f\rho[z].
\end{align*}
These operators generate the following algebra
intrinsic to
the noncommutative system:
\begin{align}\label{intrinsic}
 [Z,\:\bar Z]&=2c_1\left(S +
     f\left(2-\theta f\right)G\right),&
    [\cD,\:\bar\cD]&=-2c_1\theta G,\notag\\
 [Z,\:\bar\cD]&=2c_1\left(1-\theta f\right)G,&
    [G,\:\cD]&=\theta\cD,\notag\\
 [\cD,\:\bar Z]&=2c_1\left(1-\theta f\right)G,&
    [\bar\cD,\:G]&=\theta\bar\cD,
 \notag\\
 [Z,\:G]&=\left(1-\theta f\right)\cD,&
    [S,\:\bar\cD]&=\bar\cD,\\
 [G,\:\bar Z]&=\left(1-\theta f\right)\bar\cD,&
    [\cD,\:S]&=\cD,\notag\\\notag
 [Z,\:S]&=Z,& [S,\:\bar Z]&=\bar Z,
\end{align}
\vskip -8mm
$$
  [S,\:G]=[Z,\:\cD]=[\bar Z,\:\bar\cD]=0,
$$
where the operators $\bar\cD$, $\cD$, $G$ and $S$ in
the
chosen gauge are
\begin{align}
 \cD&=\bar z,& \bar\cD&=z,&
 G&=N +\theta\frac{c_0+c_1}{2c_1},&
 S&=-\theta^{-1}{\mop{ad}N}.
\end{align}

The algebra has the following two Casimir operators:
\begin{align}\label{casimir}
 C_1&=\left\{\bar\cD,\,\cD\right\}-2c_1G^2,
 \notag\\[-2.9mm]\\[-2.9mm]\notag
 C_2&=\left\{\theta\bar Z+(1-\theta f)\bar\cD,\,
      \theta Z+(1-\theta f)\cD\right\}
      -2c_1\left(\theta S+G\right)^2.
\end{align}
The algebra (\ref{intrinsic}) is
$\asl(2,\R)\oplus\asl(2,\R)$ ($c_1>0$) or
$\su(2)\oplus\su(2)$ $(c_1<0)$. In the chosen gauge it
can
be shown by representing the generators of the algebra
in the form ${L_z,L_{\bar z},L_N}$ and
${R_z,R_{\bar z},R_N}$, where $L_a $ and $R_a$ denote
the
left and right multiplications. In the given
representation the Casimir operators (\ref{casimir})
are
equal to the same number,
\begin{align}\label{C=}
 C_1&=C_2=\frac{\theta^2}{2c_1}
 \left(c_1^2-c_0^2\right).
\end{align}

It is necessary to note that the values $\theta f=0,
\,1,\,2$
are special from the viewpoint of the algebra
(\ref{intrinsic}). Although the algebraic content of
it
remains the same, the algebra is not generated by the
operators $Z$ and $\bZ$, i. e. it is not intrinsic any
more.
In these cases the generating elements produce
the $\asl(2,\R)$ or
$\su(2)$ only.

The operators $\cD$, $\bar\cD$ and $G$ can be treated
as
dependent covariant coordinates of a noncommutative
membrane. Indeed, the equivalent set of coordinates
\begin{align*}
 X_1&=\frac{\cD+\bar\cD}{\sqrt{8|c_1|}},&
 X_2&=i\frac{\cD-\bar\cD}{\sqrt{8|c_1|}},&
 X_3&=G
\end{align*}
obeys the condition
\begin{equation}\label{r2}
  X_1^2+X_2^2 - \varepsilon X_3^2=\varepsilon
 \theta^2\alpha\left(1-\alpha\right)=-\varepsilon r^2,
\end{equation}
where $\varepsilon=\mop{sign} c_1$ and
$\alpha=(c_1+c_0)/(2c_1)\ (c_1>0)$ or
$\alpha=-p/2$ ($c_1<0$), $p\in\Z_+$.
So, the operators $X_i$ form the set of noncommutative
covariant coordinates of the fuzzy sphere $S_\theta^2$
($c_1
<0$), or of the noncommutative hyperbolic plane $H_
\theta^2$
($c_1>0$). The operators $Z$, $\bar Z$ and $S$,
in their turn, can be interpreted as the covariant
derivatives. Note that in the
noncommutative system under consideration,
unlike the constant magnetic field case \cite{nair},
the covariant coordinates are not proportional to the
covariant derivatives.

The intrinsic algebra gives us the possibility to
define the magnetic field in a covariant way. In
general, if the derivatives form a Lie algebra,
$[\partial_i,\,\partial_j]=f_{ ij}{}^k\partial_k$, then
the corresponding gauge covariant field strength can be
defined as $iF_{ij}=[\cD_i,\,\cD_j]-f_{ ij}{}^k\cD_k$
(e. g., see Ref. \cite{nair}), where $\cD_i$ are gauge
covariant derivatives. From this definition and the
intrinsic algebra it follows that in the present case
only the ``transversal'' component ($i=3$) of the
magnetic field, $B_i= \frac 12f_i{}^{jk}F_{jk}$, does
not vanish. In the gauge (\ref {gauge}) this component
exactly coincides with the definition (\ref{B-N}). It
is interesting to note that like in the commutative
case \cite{nRS}, the magnetic field is a linear
function of the ``transversal'' coordinate.

In terms of the generators of the algebra
(\ref{intrinsic}) the higher operators of the
contracted Onsager algebra are
\begin{align*}
 Z_m&=(2c_1)^m\bigl(Z+\theta^{-1}(1-\theta f)
      (1-(1-\theta f)^{2m})\cD\bigr),\\
 B_m&=(2c_1)^m\bigl(S+\theta^{-1}
      (1-(1-\theta f)^{2m})G\bigr),
\end{align*}
where $\theta f\ne 0,1,2$. Applying this form of the
generators of the Onsager algebra to the infinite set
of the
commuting charges (\ref{jmn}), one can obtain
\begin{align*}
 \mathsf J_\lambda^k = \frac{(2c_1)^{k-1}}{
 \theta f(2-\theta f)}
  \Bigl(2\left(1-(1-\theta f)^{2k}\right)H_\lambda
   - c_1\left((1-\theta f)^2
    - (1-\theta f)^{2k}\right) S(S-\lambda)\Bigr) +
    \ldots,
\end{align*}
where dots denote a linear combination of the Casimir
operators (\ref{casimir}). This representation leads
to
the conclusion that the Hamiltonian (\ref{H}) and the
operator $S$ form a complete set of nontrivial
integrals of
the system under consideration.

\subsection{The commutative limit}

Our aim is to apply the found intrinsic algebra
of the noncommutative system with the Hamiltonian (
\ref{H})
for investigation of its spectrum. The algebraic
scheme
we shall discuss is an analogue of the Fock space
construction. It is based on the introduction of the
``ground'' vectors in the Hilbert space of the system
(\ref{H}). These vectors have special properties with
respect
to the action of the generators of the intrinsic
algebra.

First of all, let us note that in the commutative
limit
$\theta\to 0$ the algebra (\ref{intrinsic}) is
converted
into the intrinsic algebra of the system (\ref{H})
living on
the (pseudo)sphere \cite{nRS}. Indeed, with the
rescaling
\begin{align*}
  G&\to\frac G{4c_1f},&
  \cD&\to\frac\cD{2f},&
  \bar\cD&\to\frac{\bar\cD}{2f},&
  c_1\to2\beta,
\end{align*}
where the parameter $\beta$ is related to the radius
of the
(pseudo)sphere, the algebra (\ref{intrinsic}) changes
into
the intrinsic algebra appeared in Ref. \cite{nRS} in
the
limit $\theta\to 0$.

Since the commutative case is much simpler than the
noncommutative one, we shall start the discussion
of the covariant algebraic construction from it.

We introduce a vector (``ground'' state) of the
Hilbert
space which obeys the conditions
\begin{align}\label{main}
 Z\Psi^{(0)}&=\cD\alpha(G)\Psi^{(0)},&
 \bar Z\Psi^{(0)}&=\bar\cD\beta(G)\Psi^{(0)},&
 S\Psi^{(0)}&=s\Psi^{(0)}.
\end{align}
Here $\alpha(.)$ and $\beta(.)$ are some functions,
which
will be defined later, and $s\in\R$. We start with the
real
eigenvalues of the operator $S$ since no quantization
condition follows from the intrinsic algebra. Besides,
we
know \cite{nRS} that the eigenvalues do are real in
general,
since the magnetic flux gives a contribution to them
and for
a noncompact surface it is not quantized.

{}From the commutation relations for the generating
elements
$Z$ and $\bar Z$ the following constraint on the
functions
$\alpha(G)$ and $\beta(G)$ appears:
\begin{align}\label{constr0}
 \frac 12\left(G^2 - \frac{c_1^2}4\right)\varrho\prime
 (G)
  + G\varrho(G)=G - 4\beta^2s,
\end{align}
where $\varrho(G)=\alpha(G) + \beta(G)$. The solution
of the
differential equation is
\begin{align}
 \varrho (G)=1 - \frac{8\beta^2s}{G + \frac{c_1}2}
   - \frac{4\beta^2c_1m}{G^2 - \frac{c_1^2}4}.
\end{align}
The parameter $m$ is the integration constant. One
should
regard this constant as an integer since it
corresponds
to the
quantum number of the planar angular momentum of the
system
\cite{nRS}.

We will look for the eigenstates of the Hamiltonian in
the form
\begin{align}\label{eigens}
 \Psi=\sigma(G)\Psi^{(0)},
\end{align}
where we suppose that $\sigma (G)$ is a polynomial.
Then the Hamiltonian (\ref{H}) is reduced to the second
order differential operator
\begin{align}
  H'_\lambda&={}-4\beta^2\left(G^2 - \frac{c_1^2}4
  \right)
   \frac{d^2}{dG^2}
  + \left(\left(G^2-\frac{c_1^2}4\right)
  \bigl(\alpha(G)-\beta(G)\bigr)
  -8\beta^2G\right)\frac d{dG}
  \notag\\[-3mm]\label{comH}\\[-3mm]\notag&
  + G(2f+\lambda-1) + \left(G^2-\frac{c_1^2}4\right)
   \left(\alpha\prime(G)+\frac{\alpha(G)\beta(G)}{4
   \beta^2}
   \right) + const
\end{align}
acting on the function $\sigma (G)$.

Let us fix the residuary arbitrariness in the
definition of the functions $\alpha(G)$ and $\beta(G)$
requiring that the operator (\ref{comH}) is
quasi-exactly solvable. Therefore, it should be
represented in terms of the generators of
the $\asl(2,\R)$ algebra \cite{qes}
\begin{align}\label{sl2a}
 J_+&= G^2\frac d{dG} - (\mathsf N - 1)G,&
 J_0&= G\frac d{dG}- \frac{\mathsf N - 1}2,&
 J_-&= \frac d{dG},
\end{align}
where if the real parameter $\mathsf N$ is a
nonnegative integer, it defines dimensionality of the
corresponding finite-dimensional representation.
Moreover, the operator (\ref{comH}) is a quadratic
polynomial in the generators (\ref{sl2a}) since it is a
second order differential operator \cite{qes}.

Eventually one can conclude that the operator
(\ref{comH}) can be represented in the form
\begin{align}\label{slT}
 T&=c_1^2\beta^2J_-^2 - 4\beta^2J_0^2
  + a_+J_+ + a_0J_0 + a_-J_- + const,
\end{align}
where the parameters $a_\pm$, $a_0$ and $\mathsf N$
have the
following eight solutions:
\begin{align}
 \mathsf N &=\epsilon\lambda,&
 a_+ &=\epsilon, &
 a_0 &= 4\epsilon\beta^2\left(2s - \lambda\right),&
 a_- &=\frac\epsilon 4c_1\left(c_1
   + 16\beta^2(m - s)\right),
  \notag\\
 \mathsf N &=\epsilon (\lambda-m),&
 a_+ &=\epsilon , &
 a_0 &= 4\epsilon\beta^2(2s-m - \lambda),&
 a_- &= \frac\epsilon 4c_1(c_1 - 16\beta^2s),
  \notag\\
 \mathsf N &=\epsilon(\lambda - 2s), &
 a_+ &=\epsilon , &
 a_0 &=- 4\epsilon\lambda\beta^2,&
 a_- &=\frac\epsilon 4c_1(c_1 - 16\beta^2(m - s)),
  \notag\\\label{s0}
 \mathsf N &=\epsilon(\lambda + m - 2 s),&
 a_+ &=\epsilon , &
 a_0 &= 4\epsilon(m - \lambda)\beta^2,&
 a_- &= \frac\epsilon 4c_1(c_1 + 16\beta^2s),
\end{align}
where $\epsilon=\pm 1$. So, if the parameters of the
system take such values that $\mathsf N\in\Z_+$ for any
of the solutions, then the system is quasi-exactly
solvable. Any QES part of the spectrum is a
superstructure over the corresponding ``ground'' state.
Therefore, one can interpret these states as some QES
``vacua'' of the system. Here and in what follows we
will understand under the QES ``vacuum'' a state over
which one can construct a certain number of states of
the QES part of the spectrum following an
oscillator-like procedure based on an intrinsic
algebra.

The number of the $\asl(2,\R)$ representations
(\ref{slT}) coincides with the number of the solutions
found in Ref. \cite{nRS}. At the same time, it is worth
noting that the method based on the intrinsic algebra
is completely covariant since all the generators of the
algebra are covariant quantities.

We would like also to remind that this commutative
system possesses a discrete symmetry
$\Z_2\otimes\Z_2\otimes\Z_2$, which intertwines the
solutions (\ref{s0}) between themselves. The detailed
discussion of the symmetry can be found in
Ref.~\cite{nRS}.

\subsection{The noncommutative case}

To investigate the spectral problem for the
noncommutative
system with the Hamiltonian (\ref{H}) in terms of the
intrinsic algebra (\ref{intrinsic}), we introduce a
``ground'' state obeying the conditions (\ref{main})
but
with the only change
\begin{equation}\label{S=m}
 S\Psi^{(0)}=m\Psi^{(0)},\hskip 2cm
 m\in\Z.
\end{equation}
One can treat this number as an ``azimuthal'' quantum
number
since it corresponds to the existence of ``axial''
symmetry
in the system with operator $S$ as its generator.
We assume this in accordance with the representation
(\ref{modes}) of the eigenstates. Therefore, in the
noncommutative case the conserved quantity $S$ is an
analogue of the angular momentum operator. We would
like to
note that although we use formally the same conditions
(\ref{main}), one should remember that the operators $
\cD$,
$\bar\cD$ and $G$ do not commute and the ordering of
the
operators is important.

{}From the commutation relations of the generating
elements
(\ref{intrinsic}) and the definition of the ``ground''
state
(\ref{main}) it follows that the functions $\alpha(G)$
and
$\beta(G)$ are not independent but obey the constraint
\begin{multline}
\hskip 1.4mm
  2c_1G\bigl(\left(1-\theta f\right)
 \left(\alpha(G)+\beta(G)\right)
   +\theta\alpha(G)\beta(G)\bigr)
 + \bar\cD\cD\left(1-\theta f+\theta\alpha(G)\right)
  \Delta_+\beta(G)
 \\\label{constr}
 + \cD\bar\cD\bigl(1-\theta f+\theta\beta(G)\bigr)
  \Delta_-\alpha(G)=2c_1\left(m +
  f\left(2-\theta f\right)G\right).
 \hskip 1.4mm
\end{multline}
This is the noncommutative analogue of the restriction
(\ref{constr0}).

{}For investigation of the spectral problem it is
convenient
to represent the Hamiltonian as the linear in $Z$, $
\bar Z$
form:
\begin{align}
 H_\lambda&={}
 - 2\theta^{-1}\left(1-\theta f\right)
    \left(\cD\bar Z + \bar\cD Z\right)
 +  2c_1\theta^{-1}f\left(2-\theta f\right)
   G\,(G-\lambda\theta)
  \notag\\[-3mm]\label{linH}\\[-3mm]\notag&
 + 4c_1\theta^{-1}GS
 + 2c_1S\,(S - \lambda)
 - \theta^{-2}C_1\left(1-\theta f\right)^2 + \theta^{-
 2}C_2.
\end{align}
Using the definition of the ``ground'' state
(\ref{main}), the spectral problem for the Hamiltonian
in this form is reduced to that of the
finite-difference operator
\begin{eqnarray}
  {\sf H}_\lambda&=&2\left(\theta f-1\right)
  \Bigl(\cD\bar\cD\bigl(1-\theta
  f+\theta\beta(G)\bigr)
     \Delta_+\Delta_- + \bigl(\bar\cD\cD\alpha(G)
     - \cD\bar\cD\beta(G)
  \notag\\\notag&&{}
  + 2c_1\left(1-\theta f\right)G\bigr)\Delta_+
   + \theta^{-1}\left(\bar\cD\cD\alpha(G)
   + \cD\bar\cD\beta(G)\right)\Bigr)
  \notag\\\label{fdif}&&{}
  + 2c_1f\theta^{-1}\left(2-\theta f\right)
    G\,(G - \lambda\theta)
  + 4c_1m\theta^{-1}G + const
\end{eqnarray}
acting on the function $\sigma (G)$. Here the operators
$\Delta_\pm$ are defined as the difference derivatives
(\ref{dd}) but with $N$ changed for $G$. The operator
(\ref{fdif}) is a finite-difference operator in the
variable $G$ since $\bar\cD\cD$ and $\cD\bar\cD$ are
the quadratic polynomials in $G$,
\begin{align*}
  \bar\cD\cD&=c_1\left(G+\frac\theta 2\right)^2
     +\frac{\theta^2c_0}{4c_1},&
  \cD\bar\cD&=c_1\left(G-\frac\theta 2\right)^2
     +\frac{\theta^2c_0}{4c_1},
\end{align*}
as this follows from the intrinsic algebra
(\ref{intrinsic}) and the form of the Casimir operators
(\ref{casimir}), (\ref{C=}). The evident advantage of
the representation (\ref{linH}) is that the
finite-difference operator (\ref{fdif}) is linear in
the functions $\alpha(G)$ and $\beta(G)$.

Representing the functions $\alpha(G)$ and $\beta(G)$
in the form
\begin{equation*}
 \alpha(G)=\frac{\tilde\alpha(G)}{\bar\cD\cD},
 \hskip 2cm
 \beta(G)=\frac{\tilde\beta(G)}{\cD\bar\cD},
\end{equation*}
one can see that if the functions $\tilde\alpha(G)$
and $\tilde\beta(G)$ are polynomials, then the operator
(\ref{fdif}) is a finite-difference operator with
polynomial coefficient functions and vice versa.

Let us fix the form of the functions $\tilde\alpha(G)$ and
$\tilde\beta(G)$ by demanding the Hamiltonian (\ref{linH})
to be a QES operator of the form (\ref{gQES}) for some
values of the parameters. Moreover, we require that the
functions to obey the constraint (\ref{linH}). In comparison
with the commutative case the noncommutative one is
drastically different. There are four solutions for the
functions $\tilde\alpha(G)$ and $\tilde\beta(G)$ providing
the quasi exact solvability of the Hamiltonian (\ref{linH}).
The explicit form of the function is not important. The
corresponding solutions for the coefficients in the
representation (\ref{gQES}) are
\begin{gather}
 a_{1,2}=0, \qquad
 b_+^\pm ={}\mp\frac{2c_1}\theta(1-\theta
 f)^{1\pm\mu},
 \qquad
 b_0^\pm = 2c_1\bigl(2j+\mu\bigl((1\pm 1)m
  -\lambda\bigr)\bigr)(1-\theta f)^{1\mp\mu},
 \notag\\\label{s1}
 b_-^\pm ={}\pm\frac{\theta(c_1-\epsilon c_0)}{2c_1}
  \Bigl(\epsilon c_0-\bigl(4j+2\mu\bigl((1\pm\mu)m-
  \lambda
  \bigr)
  +1\bigr)c_1\Bigr)(1-\theta f)^{1\mp\mu},
\end{gather}
where $\epsilon=\pm 1$ and $\mu=\pm 1$ parametrize the
set of solutions. The main difference of these
solutions from those of the commutative case (\ref{s0})
is the absence of restrictions on the parameter $
\mathsf N =2j+1$. Therefore, the operator (\ref{fdif})
is quasi-exactly solvable since we can always choose
the parameter $\mathsf N$ to be natural.

The functions $\tilde\alpha(G)$ and $\tilde\beta(G)$
corresponding to the solution (\ref{s1}) depend on the
parameter $j$. Therefore, one can consider that it
labels the ``ground'' states as well,
$$
 \Psi^{(0)}=\Psi^{(0)}_{m,j}.
$$
Here we recovered the dependence of the state on
the quantum number $m$ (\ref{S=m}). Hence,
unlike the
commutative case, the noncommutative system admits the
infinite number of the ``ground'' states for every
value of
the quantum number $m$. These states serve as QES
``vacua''
of the model.

Like the commutative case, the noncommutative system
has a
discrete symmetry. Indeed, evidently
that the substitution
\begin{equation}\label{ds1}
  c_0\to-c_0
\end{equation}
converts the solutions (\ref{s1}) with $\epsilon=1$ to
those
with $\epsilon=-1$ and vice versa. This transformation
is
similar to the transformations of the commutative case
since it
converts one QES operator (\ref{fdif}) into another of
the
same type. Besides, in this case there is another
discrete
transformation
\begin{align}\label{ds2}
 \theta&\to-\theta,&  f&\to-f,&
 \lambda&\to-\lambda,& m&\to-m.
\end{align}
This mapping induces the following change of
the parameters (\ref{s1}) and the operators
(\ref{dslo}):
\[
 b_\alpha^\pm(\mu)\to b_\alpha^\mp(-\mu), \hskip 20mm
 J_\alpha^\pm\to J_\alpha^\mp,
\]
where $\alpha=0,\,\pm$. The transformation of the
generators is provoked by the change of the sign of the
parameter $\theta$. This follows from the definitions
of the forward and backward derivatives (\ref{dd}) and
of the generators (\ref{dslo}) themselves. In this way,
the transformation (\ref{ds2}) intertwines the QES
operators corresponding to the solutions (\ref{s1})
with $\mu=1$ and $\mu=-1$, respectively. The whole
group of the discrete symmetry consists of four
elements including the identity mapping. The
transformations (\ref{ds1}) and (\ref{ds2}) are
associated with the generators of a finite generated
Abelian group \cite{nakahara}. This discrete group is
$\Z_2\otimes\Z_2$ since all its elements are
involutive. So, like in the commutative case, this
discrete symmetry is a kind of duality.

\section{Discussion and outlook}

Let us summarize briefly the obtained results and
discuss some open problems that deserve further
attention.

The Dolan-Grady relations provide a rich algebraic
structure, which arises in numerous branches of
mathematical physics. These relations originally
appeared in the framework of various integrable models
\cite{dg,Perk,dav,gehlen,baxter}. Later it was observed
that the Dolan-Grady relations are necessary conditions
of the anomaly-free quantization \cite{d1,nsusy} of a
pseudo-classical 1D systems with the nonlinear
holomorphic supersymmetry \cite{susy-pb}. This
observation revealed the intimate connection of the
relations with the QES systems. The algebraic origin of
the construction allowed us to apply it to the
two-dimensional systems \cite{d2,nRS}. In the present
paper we demonstrated that
\begin{itemize}
\item
 The Dolan-Grady relations arise in low-dimensional
 matrix models.
\end{itemize}
The matrix models naturally lead to appearance of the
noncommutative geometry \cite{BFSS,IKKT,Taylor01}.
Using this observation as a hint, we applied the
algebraic scheme described in Ref. \cite{nsusy} to the
two-dimensional systems on a noncommutative background
subjected to an external $U(1)$ gauge interaction. We
started by postulating that the coordinates form the
generalized deformed oscillator algebra \cite{gdo}. The
Hamiltonian of the system was constructed in terms of
generating elements realized as gauge covariant
derivatives on the noncommutative manifold. We showed
that the generating elements obey the Dolan-Grady
relations only when
\begin{itemize}
\item
 The noncommutative coordinates generate $\su(2)$,
 $\asl(2,\R)$ or the Heisenberg algebras.
\end{itemize}
This allowed us to identify the resulting
noncommutative manifold, correspondingly, with the
fuzzy sphere $S^2_\theta$, the noncommutative
hyperbolic plane $H_\theta^2$ or the noncommutative
plane $\R^2_\theta$. The cases of $\su(2)$ and
$\asl(2,\R)$ were discussed in detail.

We showed that the parameter $f$ defining the intensity
of the quasi-magnetic field has the three critical
values: $f=0$, $f=\theta^{-1}$ and $f=2\theta^{-1}$.
The first case is trivial since it corresponds to the
absence of the gauge interaction ($B=0$). The second
case is analogous to the situation for the particle on
the fuzzy sphere and the noncommutative hyperbolic
plane in the presence of the constant magnetic field
$B=\theta^{-1}$ \cite{nair} (see also Ref.
\cite{noncomplane} for the case of the noncommutative
plane). For this special value the systems are
degenerate. In the third case, $f=2\theta^{-1}$, the
quasi-magnetic field vanishes while the gauge
interaction still is present in the model since the
corresponding connection is not trivial, $f\neq 0$ (see
Eqs. (\ref{gauge}), (\ref{f})). This resembles the
famous Aharonov-Bohm effect, but here its origin roots
in the noncommutativity of the configuration space.
Note that there is no analogue of such a phenomenon in
the constant magnetic field case.

By the reduction of the Hamiltonian to one-dimensional
finite-difference operators, it was showed that the
constructed noncommutative systems are quasi-exactly
solvable for some values of the coupling constant. In
the commutative limit the systems on the fuzzy sphere
and noncommutative hyperbolic plane lead to QES systems
on the corresponding Riemann surfaces \cite{nRS}.

The algebraic content of the given model was revealed
by the construction of the intrinsic algebra. Using the
notion of the intrinsic algebra, in Ref. \cite{nRS} we
proposed a general algebraic method of constructing the
covariant form of the integrals of motion. Developing
this idea, here we demonstrated how
\begin{itemize}
\item
 The spectral problem can be treated in terms of the
 intrinsic algebra.
\end{itemize}
The advantage of this approach is its complete
covariance. For example, one does not need to impose
any gauge conditions since the intrinsic algebra
operates with covariant quantities. The application of
this method allowed us to show that the noncommutative
systems under consideration are quasi-exactly solvable
for any value of the coupling constant.

In the noncommutative case the reduction of the
Hamiltonian and the application of intrinsic algebra to
the spectral problem lead to some finite-difference
equations. To investigate their quasi exact
solvability,
\begin{itemize}
\item
 A deformation of the $\asl(2,\R)$ scheme for
 finite-difference QES systems was proposed.
\end{itemize}
In this approach the $\asl(2,\R)$ algebra is changed
for its nonlinear analogue (\ref{dsla0}). We showed
that the deformed scheme is not equivalent to the
approach based on the usual $\asl(2,\R)$ algebra
\cite{smirnov} if one requires for finite-difference
QES operators to be the polynomials in the generators
of the corresponding algebras.

Surprisingly enough, but the system (\ref{H}) in a
non-constant magnetic field on the noncommutative plane
(the coordinates form the Heisenberg algebra) is more
complicated than the systems on the fuzzy sphere and
the noncommutative hyperbolic plane investigated in
this paper. Therefore, it is of interest to investigate
this system and verify if it inherits such a property
as the quasi exact solvability. Furthermore, it would
be interesting to apply the observed analogy between
the matrix equations and the parafermionic trilinear
commutation relations \cite{kam} to construct new solutions.
Such a consideration will be presented elsewhere
\cite{pmatrix}.

\vskip 0.5cm
{\bf Acknowledgements}
\vskip 5mm

We thank W. Orrick and J. Perk for useful comments.
M.P. is also grateful to A. Turbiner for valuable
discussions and kind hospitality during his visit to
Universidad Nacional Autonoma de Mexico, where part of
this work was done. The work was supported by the
grants 1010073 and 3000006 from FONDECYT (Chile) and by
DICYT (USACH).

\appendix
\section{Deformed $\asl(2,\R)$ scheme}

Here we discuss a deformation of the $\asl(2,\R)$
scheme related to one-dimensional finite-difference
operators, and compare the approaches based on
the deformed algebra and on the usual one.

In Ref. \cite{smirnov}
Smirnov and Turbiner proposed the approach to
finite-difference QES equations based on
the following representation of the Heisenberg
algebra:
\begin{equation}\label{fdHa}
 a=\Delta_+,\hskip 2cm
 b=x(1-\theta\Delta_-),
\end{equation}
where $x$ is a real variable and $\Delta_\pm$ are
the forward
and
backward difference operators (\ref{dd}) with respect
to
this
variable. These operators obey the standard
commutation
relation, $[a,b]=1$.

Like in the standard approach to 1D differential QES
equations, an operator for the corresponding spectral
problem is supposed to be a polynomial in the $\asl(2,
\R)$
generators
\begin{align}\label{sl2}
 J_+&=b^2a-(\mathsf N-1)b,&
 J_0&=ba - \frac{\mathsf N-1}2,&
 J_-&=a.
\end{align}
When $\mathsf N\in\N$, the space of polynomials
$\cP_{\mathsf N-1}(x)$ is invariant
with respect to the action of these operators. Hence,
any
polynomial in the generators (\ref{sl2}) is an
invariant
operator on this space as well, i. e. it is a QES
operator.

We are interested in the spectral problem for a
three-point finite-difference equation,
\begin{equation}\label{3p}
 A(x)\Psi(x+\theta) + B(x)\Psi(x) + C(x)\Psi(x-\theta)
 =E\Psi(x).
\end{equation}
Ref. \cite{smirnov} claims that all such QES
equations with finite number of polynomial
eigenfunctions
are classified via the cubic polynomial element of the
universal enveloping algebra of $\asl(2,\R)$ taken in
the
representation (\ref{fdHa}):
\begin{equation}\label{T}
 {\sf T}=A_+\left(J_+ + \theta J_0^2\right)
  + A_1J_0^2\left(\theta J_- + 1\right)
  + A_2J_0J_- + A_3 J_0 + A_4J_- + const,
\end{equation}
where the coefficients are arbitrary real parameters.
Below
we shall show that this operator is not the most
general
among QES operators corresponding to the three-point
problem
(\ref{3p}).

Let us introduce the finite-difference operators which
are
linear in the finite-difference derivatives,
\begin{align}
 \label{dslo}
 J_+^{\pm}&=x^2\Delta_\pm -2jx,&
 J_0^{\pm}&=x\Delta_\mp-j,&
 J_-^{\pm}&=\Delta_\mp.
\end{align}
The half-integer parameter $j$ ($2j\in\Z_+$) is an
analogue of ``spin'' in the usual $\asl(2,\R)$ scheme
\cite{qes}. These linearly independent operators are
finite-difference analogues of the $\asl(2,\R)$
generators, which can be reproduced in the commutative
limit, $\theta\to 0$. The operators (\ref{dslo}) have
another useful property of the $\asl(2,\R)$ generators.
They have the invariant finite-dimensional subspace,
which is the space of polynomials $\cP_{2j}(x)$,
$2j\in\Z_+$, $\mop{dim}\cP_{2j}(x)=2j+1=\mathsf N$.
This allows us to use the generators (\ref{dslo}) for
the construction of finite-difference QES operators in
the manner analogous to the continuous case. The
exhaustive investigation of this construction lies out
of the scope of the paper and here we will discuss just
some aspects of it.

The both sets $J_\mu^+$ and $J_\mu^-$, $\mu=0,\pm$,
form the nonlinear subalgebras of the form
\begin{align}
 \left[J_-^{\pm},\,J_+^{\pm}\right]&=
   2J_0^{\pm}\pm\theta(2j-1)J_-^{\pm},
  \notag\\\label{dsla0}
 \left[J_0^{\pm},\,J_+^{\pm}\right]&=J^\pm_+
   \mp\theta\left(2J^\pm_+J^\pm_-
    + (2j+1)(J^\pm_0+j)\right),
  \\\notag
 \left[J_-^\pm,\,J_0^\pm\right]&=
   J_-^\pm T_\mp,
\end{align}
where the operators
\begin{equation}\label{T+-}
 T_\pm=1\pm\theta J_-^\mp
\end{equation}
are the discrete forward and backward translation
generators, $T_\pm\psi(x)=\psi(x\pm\theta)$.
The intertwining commutation relations between
the sets
$J_\mu^+$ and $J_\mu^-$ are
\begin{align}
 \left[J_-^\mp,\,J_+^\pm\right]&=
   \left(2J_0^\mp\pm\theta J_-^\mp\right)T_\pm,&
  \left[J_+^+,\,J_+^-\right]&=\theta
    \left(2J_+^+J_0^+ - J_+^+ + (2j+1)J_+^-\right),
  \notag\\
 \left[J_0^\mp,\,J_+^\pm\right]&=
   \left(J_+^\pm\pm\theta(J_0^\mp+j)\right)T_\pm,&
  \left[J_0^+,\,J_0^-\right]&=J_0^+ - J_0^-,
  \notag\\\label{dsla1}
 \left[J_-^{\mp},\,J_0^{\pm}\right]&=J_-^{\mp},&
  \left[J_-^+,\,J_-^-\right]&=0.
\end{align}
It is obvious that in the commutative limit, $\theta
\to 0$,
the nonlinear commutation relations (\ref{dsla0}),
(\ref{dsla1}) tend to the usual $\asl(2,\R)$ algebra.
Furthermore, for the realization (\ref{dslo}) of
the nonlinear algebra the following quadratic
identities
hold:
\begin{align}
 \notag&
  J^\pm_0J^\pm_0 - J^\mp_+J^\pm_- - J^\pm_0
  \pm \theta\left(J^\pm_0+j\right)J^\pm_- - j(j+1)=0,
 \\\notag&
 \theta\left(J^+_0+j\right)\left(J^-_0+j-1\right)
 + J_+^- - J_+^+=0,
 \\\notag&
 J_+^+J_0^+ - J_+^-J_0^-
   - (j+1)\left(J_+^+ - J_+^-\right)=0,
 \\\label{2constr}&
 J_+^-T_+ - J_+^+ + 2j\,\theta\left(J_0^- +
 j\right)=0,
 \\\notag&
 \left(J^-_0+j\right)T_- - J^+_0-j=0,
 \\\notag&
 J_-^-T_- - J_-^+  = 0.
\end{align}
Taking into account these identities and the
commutation
relations (\ref{dsla0}), (\ref{dsla1}), the most
general
three-point operator (\ref{3p}), quadratic in the
generators (\ref{dslo}), can be written as
\begin{eqnarray}
 H_{QES}&=& a_1J_+^+J_+^- + a_2J_+^-J_0^-
   + b_+^+J_+^+ + b_0^+J_0^+ + b_-^+J_-^+
   \notag\\[-3mm]\label{gQES}\\[-3mm]\notag&&{}
   + b_+^-J_+^- + b_0^-J_0^- + b_-^-J_-^- + const.
\end{eqnarray}
By the construction this operator is quasi-exactly
solvable.

The generators of the $\asl(2,\R)$ algebra in the
representation (\ref{sl2}) can be represented as
polynomials
in the operators (\ref{dslo}),
\begin{align}\label{0to1}
 J_+ &= \left(J_+^- - \theta(J_0^++j)\right)T_-,&
 J_0 &= J_0^+, &
 J_- &=J_-^-.
\end{align}
Hence, the QES scheme based on the $\asl(2,\R)$
generators (\ref{sl2}) is included into that based on
the operators (\ref{dslo}). On the other hand, the
operators (\ref{dslo}) can also be represented in terms
of (\ref{sl2}) via the two last equalities in
(\ref{0to1}) and
\begin{align}\label{1to0}
 J_+^+ &=J_+ + \theta\left(J_0+j\right)
 \bigl(\theta\left(J_0+j\right)J_- + 2J_0 - 1\bigr),&
 J_-^+ &=J_-\left(1 + \theta J_-\right)^{-1},
 \notag\\[-3mm]\\[-3mm]\notag
 J_+^- &=J_+T_+ + \theta(J_0 + j),&
 J_0^- &=J_0 + \theta\left(J_0 + j\right)J_-.\hskip 15
 mm
\end{align}
According to these relations the shift operator $T_+$
(\ref{T+-}) is linear in the generator $J_-$. As we
see,
the correspondence is not of a polynomial form for the
operator $J_-^+$. Therefore, the $\asl(2,\R)$ scheme
(\ref{sl2}) is not equivalent to that based on the
operators
(\ref{dslo}) if one restricts QES operators to a
polynomial
in the generators of the  (\ref{sl2}) and (\ref{dslo})
forms.
One can also notice that the 5-parametric (up to
overall
shift) QES operator (\ref{T}) is not the most general
QES
operator in the framework of $\asl(2,\R)$ scheme.
Indeed,
for the case $b_3=0$ the QES operator (\ref{gQES}) can
be
represented as a 5-th degree polynomial in the
generators
(\ref{sl2}) with 7 arbitrary real parameters.

In addition, we note that if one imposes the following
restrictions on the parameters of the QES operator
(\ref{gQES})
\begin{align}
 a_1 &= 0,& a_2 &= 0,& b_+^- &= - b_+^+,
\end{align}
then it becomes exactly solvable.


\end{document}